\documentclass[twocolumn,prl]{revtex4-1}
\usepackage{graphicx}
\usepackage{float}
\usepackage{subcaption}
\begin{document}
\title{\textbf{Application of Association rule analysis to study the evolution of halos in Cosmological N-Body simulations}}
\author{B Hareesh Gautham, Dr Rahul Nigam}
\affiliation{Birla Institute of Technology and Science-Pilani, Hyderabad Campus}

\begin{abstract}
Merger trees track the evolution of halos across multiple snapshots.  They assign for halos of a particular snapshot, the set of halos from previous snapshots they possibly originated from. In this work, Association rule analysis a well known technique from data mining has been used to build halo merger trees. Association rule analysis tries to find associations between different halos(in same as well as in snapshots) using the particle IDs of the particles which the halos are made of. Associations are expressed in the form of association rules. Merger trees are one of the several useful results one can obtain from the output of association rule analysis. Other results including halo substructure and halo splitting can also be extracted. Each type of output to be extracted from the association rule analysis output correspond to a pattern in association rules. Merger trees were formed and tested using the above technique. Dark matter simulations were run using Gadget-2 for $128^3$ particles. Halos were extracted from the simulation snapshots using Amiga Halo Finder. Halo accretion history was plotted and compared against those formed using AHF merger tree builder. 
\end{abstract}

\maketitle
\section{Introduction}
N-Body simulations offer valuable insights into non-linear structure formation. These simulations track N-particles(hence the name) over multiple time steps. They do this by (recursively) calculating gravitational force on each particle using their positions and then incrementing their positions and velocities in each time step.  To extract useful results from simulation data, tools like halo finders and merger tree builders are needed. Halo finders find gravitationally bound objects in the simulation whereas Merger trees tries to track halos across multiple snapshots. It tries to assign for halos of a particular  snapshot, the set of halos from previous snapshots it came from.

Following work tries to solve the problem of building merger trees using a well known problem in data mining, namely association rule analysis. Merger trees in the past(see \cite{mt1},\cite{mt2}) have been built by identifying halos in different snapshots which has the most number of same particles. Association rule analysis does that, but will also give additional information including halo substructure. Particles which were missed by the halo finder in one snapshot only to appear in later halos can also be tracked using this method.  Association rule analysis is a well studied and important problem in data mining. Since data mining deals with large amounts of data, any standard implementation of association rule analysis tool should be able to handle the amount of data generated by N-Body simulations.

\section{Association rule analysis}
Association Rule analysis searches for associations among various items in a data set. An example would be transactions at a supermarket.Each transaction is given by a set of items. For example a transaction could be $\{bread, butter\}$ indicating that someone bought bread and butter together.  Association Rules analysis tries to find correlations between items by analysing all transactions. Results of such analysis gives association rules of the form: 
\begin{equation}
 \{Bread\} \rightarrow \{milk\} (support =0.5, confidence = 0.8 ).
 \end{equation}
This rule indicates a strong association between bread and milk. Support of 0.5 indicates that in half of all the transactions bread and milk both occur. It is an indicator of how relevant the rule is. Confidence of 0.8 indicates that on an average, for every 10 occurrences of bread, milk occurs with bread for 8 times. This is an indicator of how strong the association is. Following is a formal description of the above concepts.

Let $I$ be the set of all items in the data set. In the above example, it would be the set of all items sold in the supermarket. Any subset of $I$ is called item set. A transaction($t$) is a subset of $I$, containing one or more items. The whole data set($T$) will contain many transactions. Association analysis searches for associations among items in $I$ using the set of transactions $T$.  Number of times an item set occurs in the whole data set is called it's support count. It should be noted that for an item set to occur in a data set, it has to be a subset of some transaction. For an item set A, it's support is given by:
\begin{equation}
\sigma(A)=\mid\{ t_i \mid A\subset t_i, t_i \subset T\}\mid
\end{equation}
Output of association analysis would be a set of rules of the form:
\begin{equation}
A \rightarrow B \quad \textrm{where},  A \subset I, B\subset I \quad\textrm{and} \quad A \cap B=\phi
\end{equation}
Which essentially says that $A$ and $B$ are disjoint item sets and $A$ is associated with $B$.
For each association rule, two values namely support and confidence can be calculated. These are indicators of how good the rule is. Support of a rule is defined as,
\begin{equation}
s(A\rightarrow B)=\frac{\sigma(A\cup B)}{\mid T \mid}
\end{equation}
Support essentially tells us how relevant the rule is. A rule with low support is of lower significance than that of a rule with higher support. A rule of low support could be just a matter of chance. Confidence of a rule is given by,
\begin{equation}
c(A\rightarrow B)=\frac{\sigma(A\cup B)}{\sigma(A)} \label{conf}
\end{equation}
Confidence tells us how strong the association is. When association rule analysis is used on a data set, thresholds on confidence and support are also given as inputs.  All rules which have support and confidence greater than these threshold values are included in the results. It should be noted that association rules DO NOT imply causality. They merely point out correlations between items by analysing the data set. Also note that the rule, $A\rightarrow B$ is different from $B\rightarrow A$. The first one indicates that all items in A are correlated with the items in B. Not the other way. Second rule is just the opposite of first. This comes from the definition of confidence in equation \ref{conf}.
\subsection{An Example}
Following is a set of transactions.
\\
\begin{tabular}{|l|l|l|l|l|l|l|}
\hline
	TID & Item 1 & Item 2 & Item 3 & Item 4 & Item 5 & Item 6\\
\hline
	1 &rice&wheat&bread&butter&&\\
\hline
	2 & oil&wheat&milk&&&\\
\hline
	3 & milk&chips&bread&&&\\
\hline
	4 & bread&rice&wheat&butter&oil&milk\\
\hline
	5 & milk&chips&rice&butter&biscuits&bread\\
\hline
	6 & milk&ice-cream&wheat&bread&&\\
\hline
	7 & oil&wheat&rice&butter&milk&bread\\
\hline
	8 & rice&wheat&&&&\\
\hline
	9 & butter&milk&ice-cream&bread&&\\
\hline
	10 &chips&bread&ice-cream&&&\\
\hline
\end{tabular}
It contains following 9 items,
\begin{eqnarray*}
I&=& \{rice,wheat,bread,butter,oil, \\
&& milk,chip,biscuits,ice cream\}
\end{eqnarray*}
It contains a total of 10 transactions. When association rule mining is applied with thresholds of support = $50\%$ and confidence = $80\%$, following rules would be obtained:
\begin{eqnarray*}
\{milk\} &\rightarrow &\{bread\} ( s=60\%,c= 85.71\%) \\
 \{rice\} &\rightarrow &\{bread\} (s=40\%,c= 80\%) \\
 \{rice\} &\rightarrow &\{wheat\} (s=40\%,c= 80\%) \\
 \{butter\} &\rightarrow &\{bread\} (s=50\%,c= 100\%) \\
 \{butter,bread\} &\rightarrow &\{milk\} (s=40\%,c= 80\%) \\
 \{butter\} &\rightarrow & \{milk\} (s=40\%,c= 80\%) \\
 \{butter,bread\} &\rightarrow & \{rice\} (s=40\%,c= 80\%) \\
 \{butter\} &\rightarrow &\{rice\} (s=40\%,c= 80\%) \\
 \{rice\} &\rightarrow &\{butter\} (s=40\%,c= 80\%) \\
\end{eqnarray*}
Since the data set is small one can check if the rules satisfy the threshold conditions.
\section{Applying association rule analysis to N-Body simulations}
\begin{figure}[h]
\centering
\includegraphics[scale=0.4]{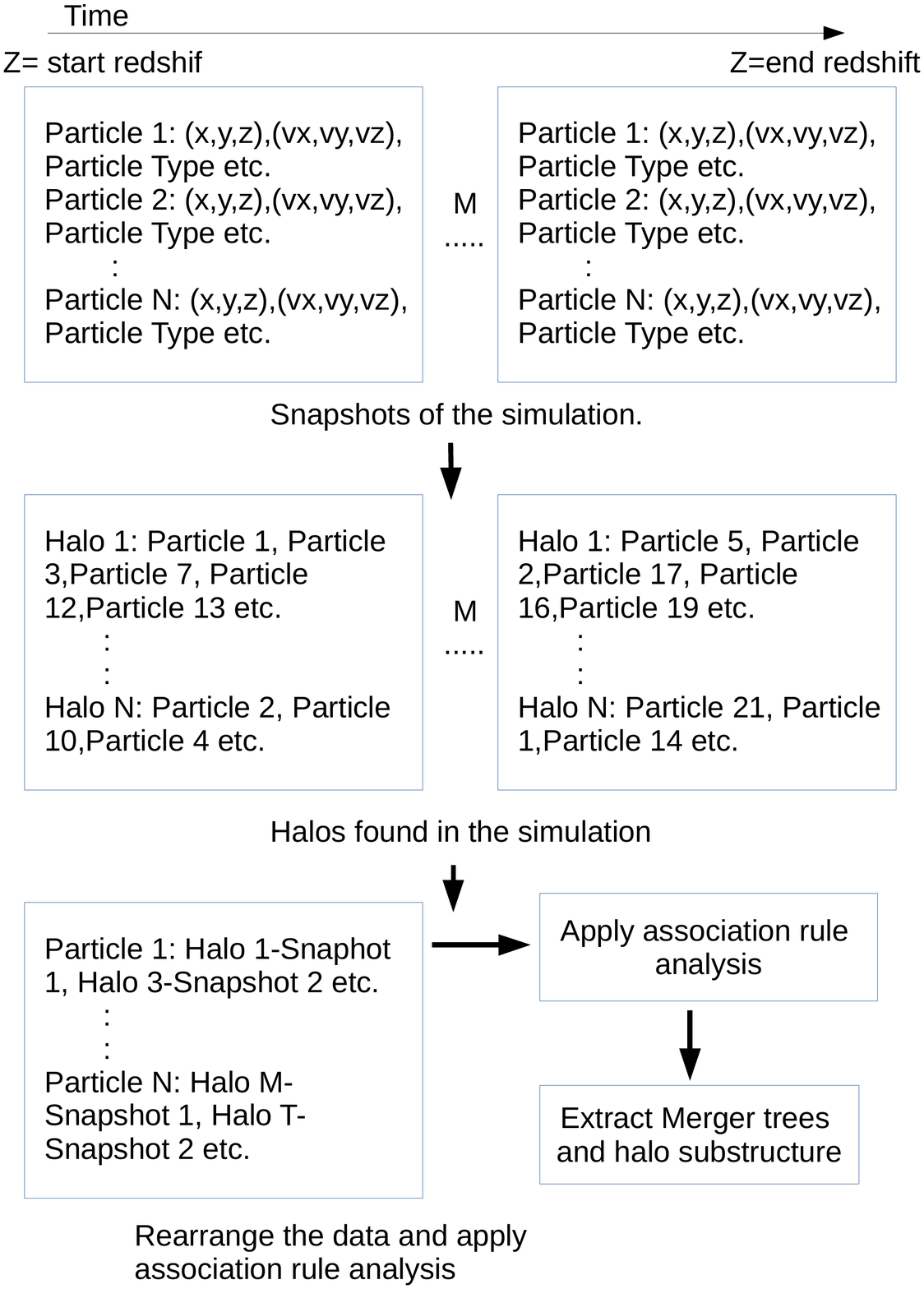}
\caption{Flowchart for getting merger trees using association rule analysis.} 
\label{fig:fchart}
\end{figure}

Association rule analysis can be applied to generate halo merger trees. Following are the steps which need to be followed to accomplish that:
\begin{enumerate}
\item Run the simulation and get the snapshots.
\item Get halos in each snapshot by running a halofinder. Particles across multiple snapshots are tracked using their particle IDs. Hence, to build a merger tree, particle IDs of all the particles in every halo found by the halo finder should also be recorded. Each halo is an item in the context of association rule analysis.
\item For each particle form a list of all the halos it has been a part of in the past. These lists will be the transactions for association rule analysis.
\item Run association rule analysis over this list to get rules which relate halos at different snapshots. These rules are nothing but a merger tree.
\end{enumerate}
Step one will produce the snapshots. They are nothing but the positions and velocities of all the particles tracked and recorded at various times specified during the simulation. Halofinders are tools used to find gravitationally bound objects in the simulation. Two popular methods to do this are Friend of friend and spherical overdensities. Halofinders usually provide various properties of the halos. These include mass, position, velocity, angular velocity among others. Since the halos are tracked across multiple snapshots using particle IDs, IDs of all the particles in each halo should be recorded. Each halo identified by the halofinder (no matter which snapshot it belongs to) will correspond to a item when association rule analysis is applied. The data obtained by halofinders in the form of halos and the IDs of the particles which constitute them, can also be written as a list of halos which each particle has been in it's entire history. Data in this form can be used as input for association rule analysis. Figure \ref{fig:fchart} shows these steps in a flow chart.

Following table shows what transactions in this context look like. They are nothing but halo history of each particle. It should be noted that not all particles will have a corresponding halo in each snapshot. Hence all transactions won't be of same length.
\begin{tabular}{|p{15mm}|p{25mm}|p{3mm}|p{25mm}|}
\hline
	Particle ID & Halo No.\newline(Snapshot No.)&....& Halo No.\newline(Snapshot No.)\\
	\it{(TID)} & \it{(Item)} & ... & \it{(Item)}\\	
\hline
	1 & Halo 1\newline(Snapshot 1) &....&  Halo M\newline(Snapshot P)\\
\hline
	2 & Halo 5\newline(Snapshot 2) &....& Halo M1\newline(Snapshot P1)\\
\hline
	: & : &:& :\\
\hline
	N-1 & Halo 10\newline(Snapshot 5) &....& - \\
\hline
	N & Halo 15\newline(Snapshot 3) &....& Halo M2\newline(Snapshot P2)\\
\hline
\end{tabular}

The corresponding rules would look like:
\begin{eqnarray*}
\{ \text{Halo 1(Snapshot 1)} \}&\rightarrow&\{ \text{Halo 2(Snapshot 2)} \} \\
\{ \text{Halo 10(Snapshot 1)} \}&\rightarrow&\{ \text{Halo 8(Snapshot 1)} \} \\
\{ \text{Halo 3(Snapshot 1)} \}&\rightarrow&\{ \text{Halo 6(Snapshot 2)} \} \\
\{ \text{Halo 3(Snapshot 1)} \}&\rightarrow&\{ \text{Halo 5(Snapshot 2)} \}
\end{eqnarray*}
Each of these rules would have corresponding support and confidence, which would obviously depend on the simulation and halo finder. Note than rule one is indicating that halo 1 of snapshot 1 becomes halo 2 of snapshot 2. Rule 2 indicates presence of a substructure. It says that Halo 10 of snapshot 1, is associated with halo 8 of the same snapshot. This would mean that Halo 10 is a subhalo of Halo 8 if confidence is $100\%$. Rules 3 and 5 indicate that halo 3 of snapshot 1 is split into two halo(Halo 6 and Halo 5) in snapshot 2. The amount of spilt is given by the corresponding confidence of the rules. The above examples are just for illustrative purpose, rules for real data can be way more complex.
\onecolumngrid
\begin{center}
\begin{figure}
\centering
\includegraphics[scale=0.38]{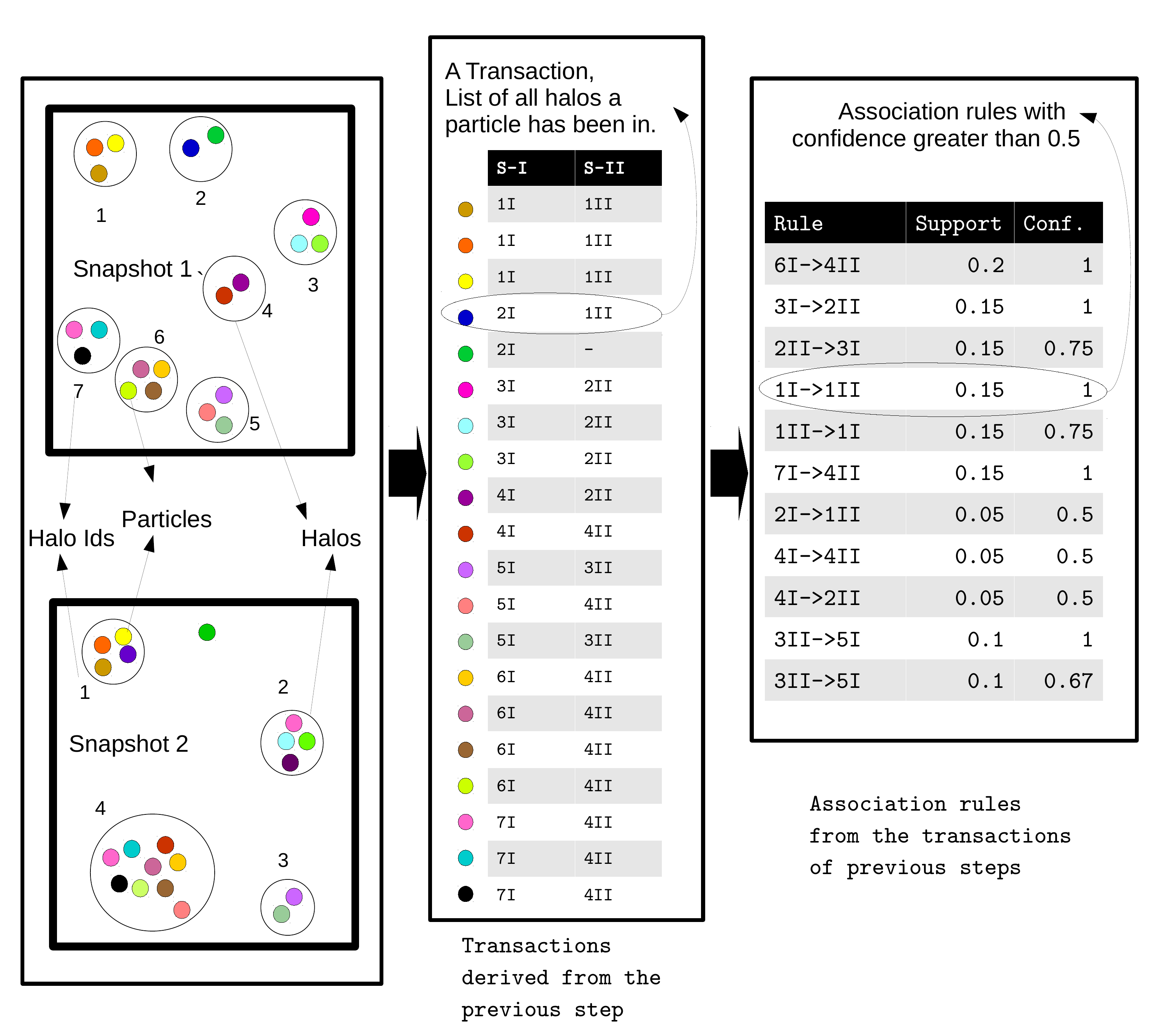}
\caption{Figure shows the application of association rule analysis to build merger trees. Each particle is assigned a color which is like it's particle ID. Two snapshots are shown in the first panel with different arrangement of particles, hence different halo structures. Halos are also identified and given IDs. This data is converted into a set of transactions in the following step. This is nothing but a record of all the halos each particle has been in. Association rule analysis is then applied to this set of transactions to get association rules and hence the merger tree.}
\label{fig:hist}
\end{figure}
\end{center}
\onecolumngrid
\twocolumngrid
\section{Implementing association rule analysis and time complexity}
There are many ways to implement association rule analysis. Brute force method is to generate all of them and then prune them by calculating their support and confidence. This could be very expensive. A data set of d different items, would have $3^d -2^{d+1} + 1 $ different association rules. After generating these rules calculating their support and confidence would require iterating over all transactions. It would be very expensive to do that. Most of the algorithms split the problem of find association rules into two sub-problems: finding frequent item sets(item sets with support greater than the specified support threshold) and generating rules from the frequent item sets.
An important method to generate frequent item sets is the apriori algorithm. It is based on the principle that if an item set is not frequent then all of it's supersets are not frequent. Similarly, if an item set is frequent then so are it's subsets. The algorithm starts with all the 1-item sets(as in item sets with one item). item sets which are not frequent are pruned. Those which are frequent are used to form 2-item sets. Of these item sets, those which do not satisfy the threshold requirement are pruned. This process is repeated till there are no more item sets left. Generating frequent items efficiently using apriori is not trivial, as there are other important decisions to be made which can influence the performance. These include how to calculate support efficiently, how to generate candidate item sets(to generate potential k-item sets from frequent (k-1)-item sets) etc. For more details see \cite{dmbook}. 

An alternate approach to generating frequent item sets is the fp tree algorithm. It is based on apriori principle but does not generate item sets to prune them later. It uses a prefix tree called fp tree to store the data in a compact form and then traverses the tree to generate frequent item sets. This work uses f-p trees to generate frequent item sets. For more details see \cite{dmbook} and \cite{fptpaper}.

Once the item sets are formed, rules need to be generated. An interesting observation can reduce the running time of this step. If a rule $X\rightarrow Y-X$,($X$ and $Y$ are item sets) does not satisfy the confidence threshold, then neither does the rule, $X'\rightarrow Y-X'$ where $X' \subset X$. This is because the support count of a set is always equal to or lesser than that of it's subsets. 

Since the algorithms mentioned above are based on support based and confidence based pruning, the running time will depend a lot on the specific data set used as well as the values of confidence and support thresholds. The running time can be reduced by reducing the number of items. This can be done by analysing two snapshots at a time. There is also a computational cost in parsing the output of association rule analysis to identify rules which are relevant to the context of merger trees.

\begin{figure*}[h]
    \centering
    \begin{subfigure}[b]{0.475\textwidth}
        \centering
        \includegraphics[width=\textwidth]{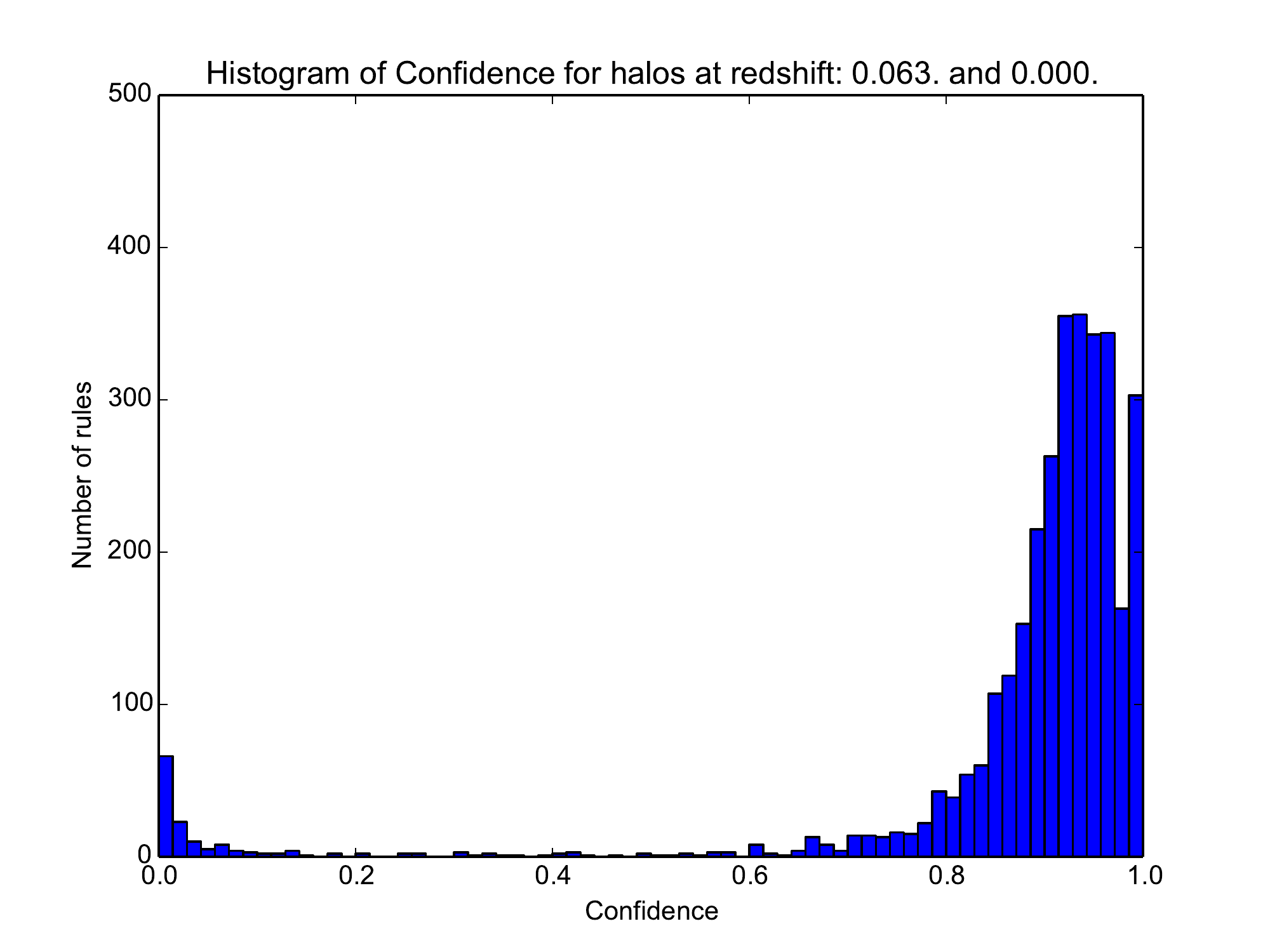}
        \caption[]%
        {{\small}}
    \end{subfigure}
    \hfill
    \begin{subfigure}[b]{0.475\textwidth}  
        \centering 
        \includegraphics[width=\textwidth]{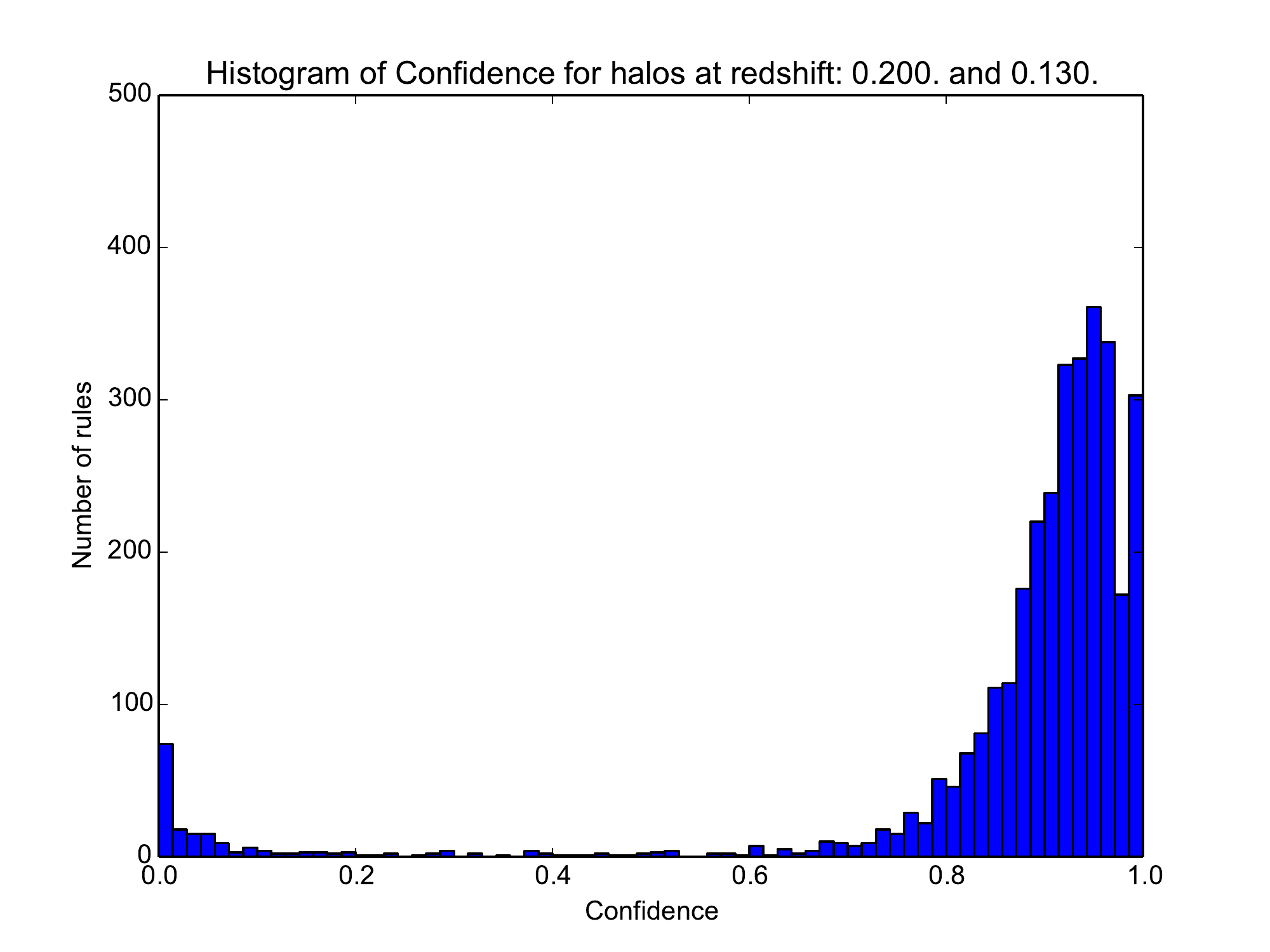}
        \caption[]%
        {{\small}}
    \end{subfigure}
    \vskip\baselineskip
    \begin{subfigure}[b]{0.475\textwidth}   
        \centering 
        \includegraphics[width=\textwidth]{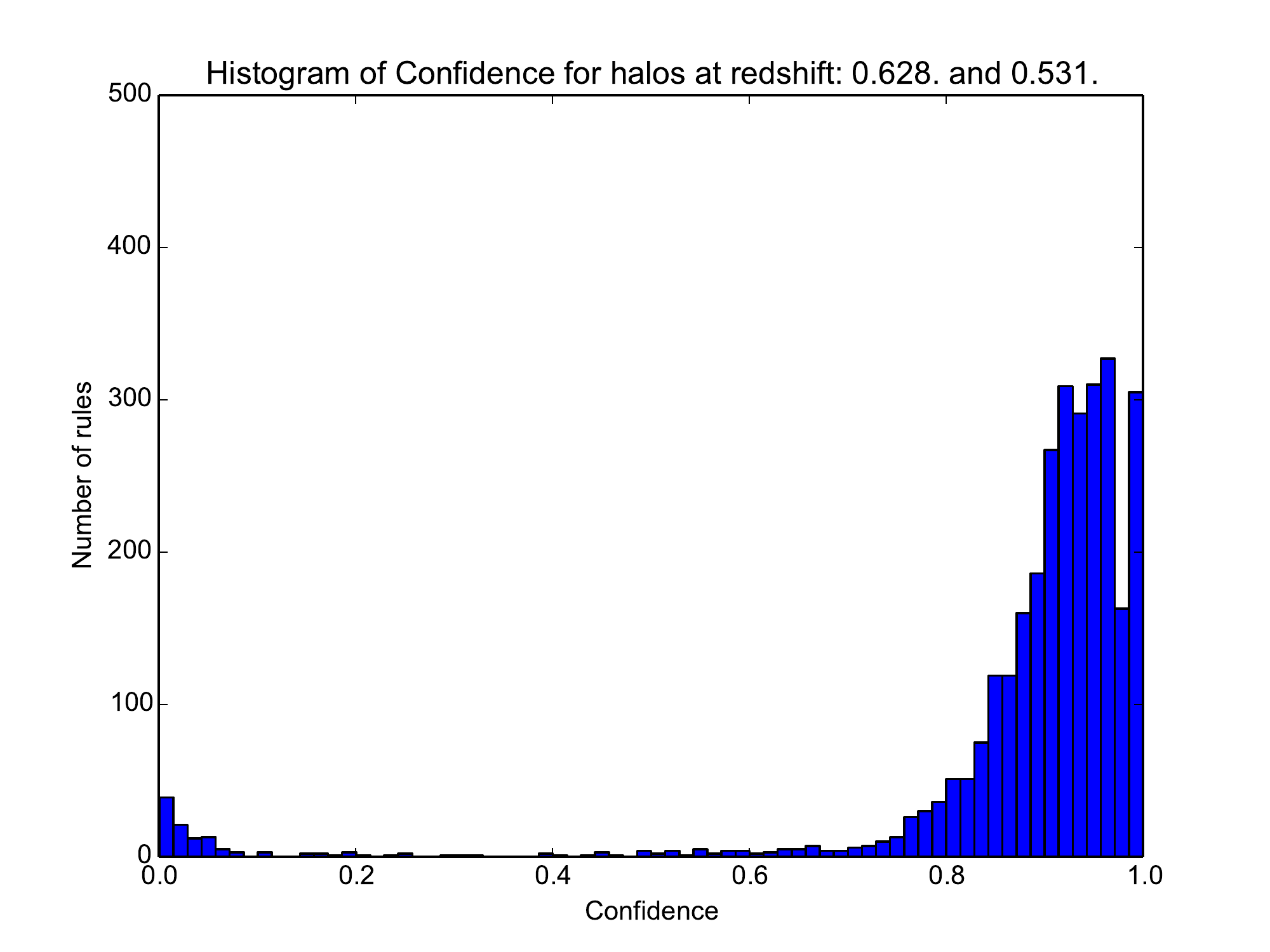}
        \caption[]%
        {{\small}}
    \end{subfigure}
    \quad
    \begin{subfigure}[b]{0.475\textwidth}   
        \centering 
        \includegraphics[width=\textwidth]{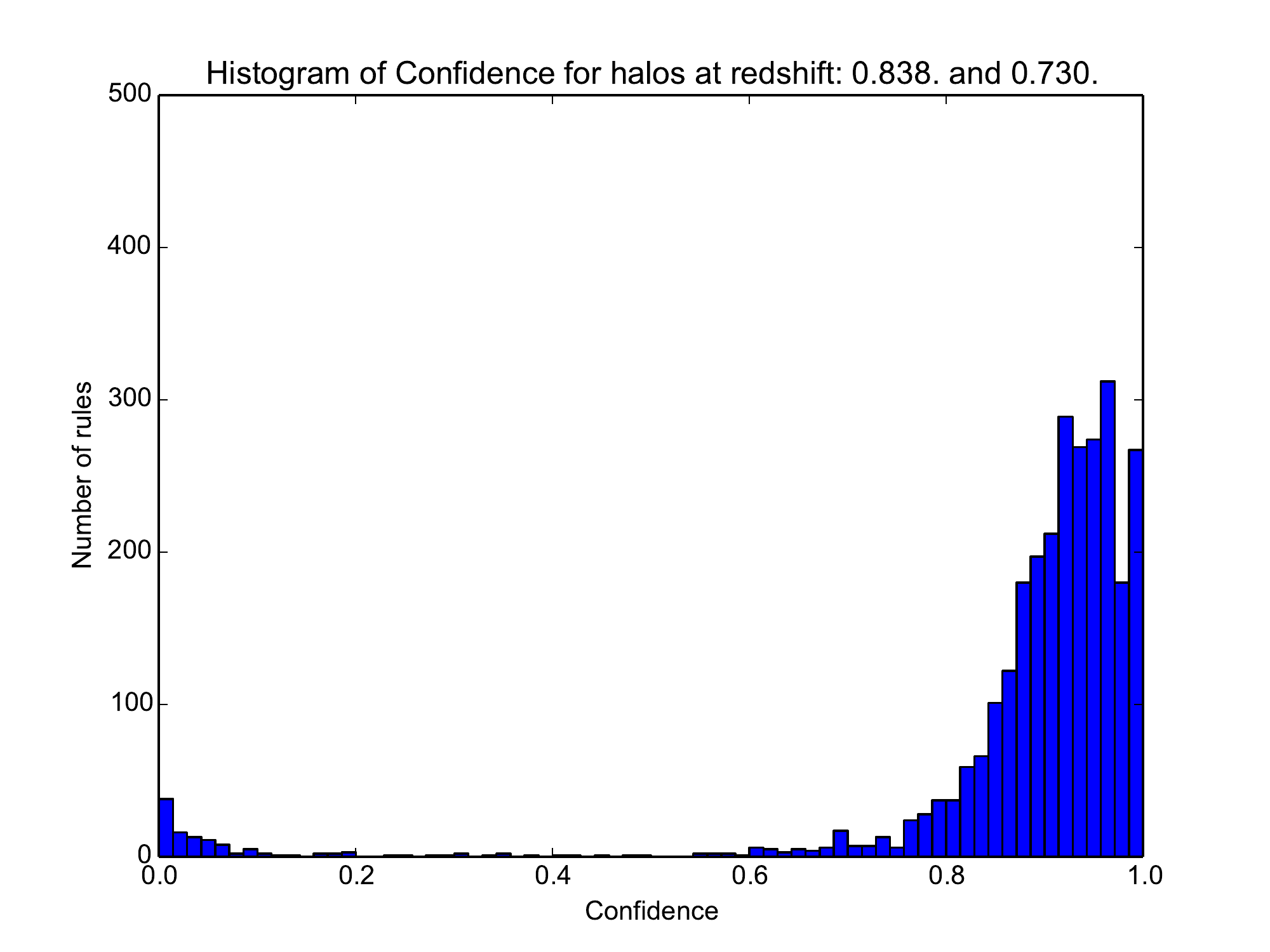}
        \caption[]%
        {{\small}}
    \end{subfigure}
    \begin{subfigure}[b]{0.475\textwidth}   
        \centering 
        \includegraphics[width=\textwidth]{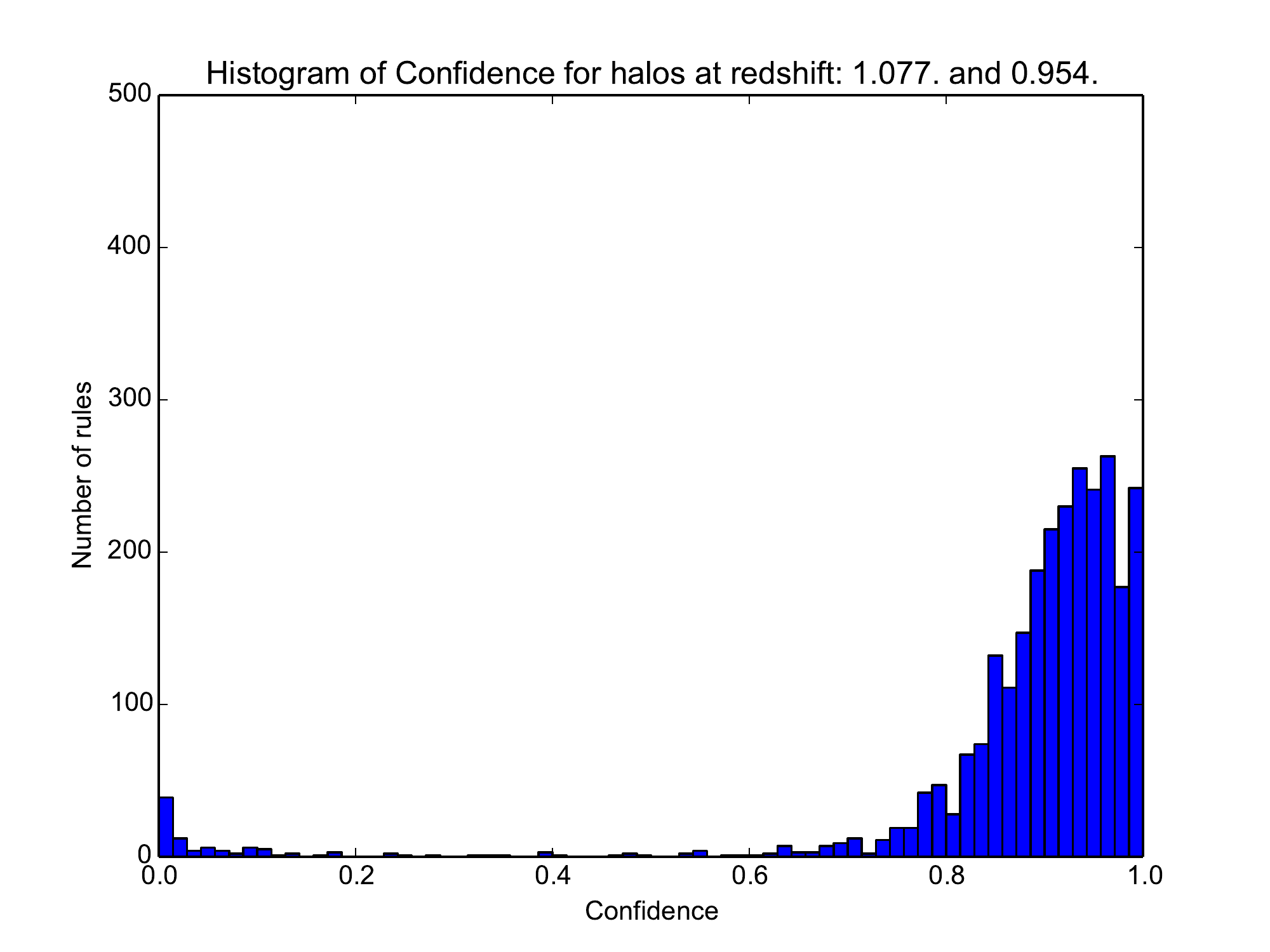}
        \caption[]%
        {{\small}}
    \end{subfigure}
    \begin{subfigure}[b]{0.475\textwidth}   
        \centering 
        \includegraphics[width=\textwidth]{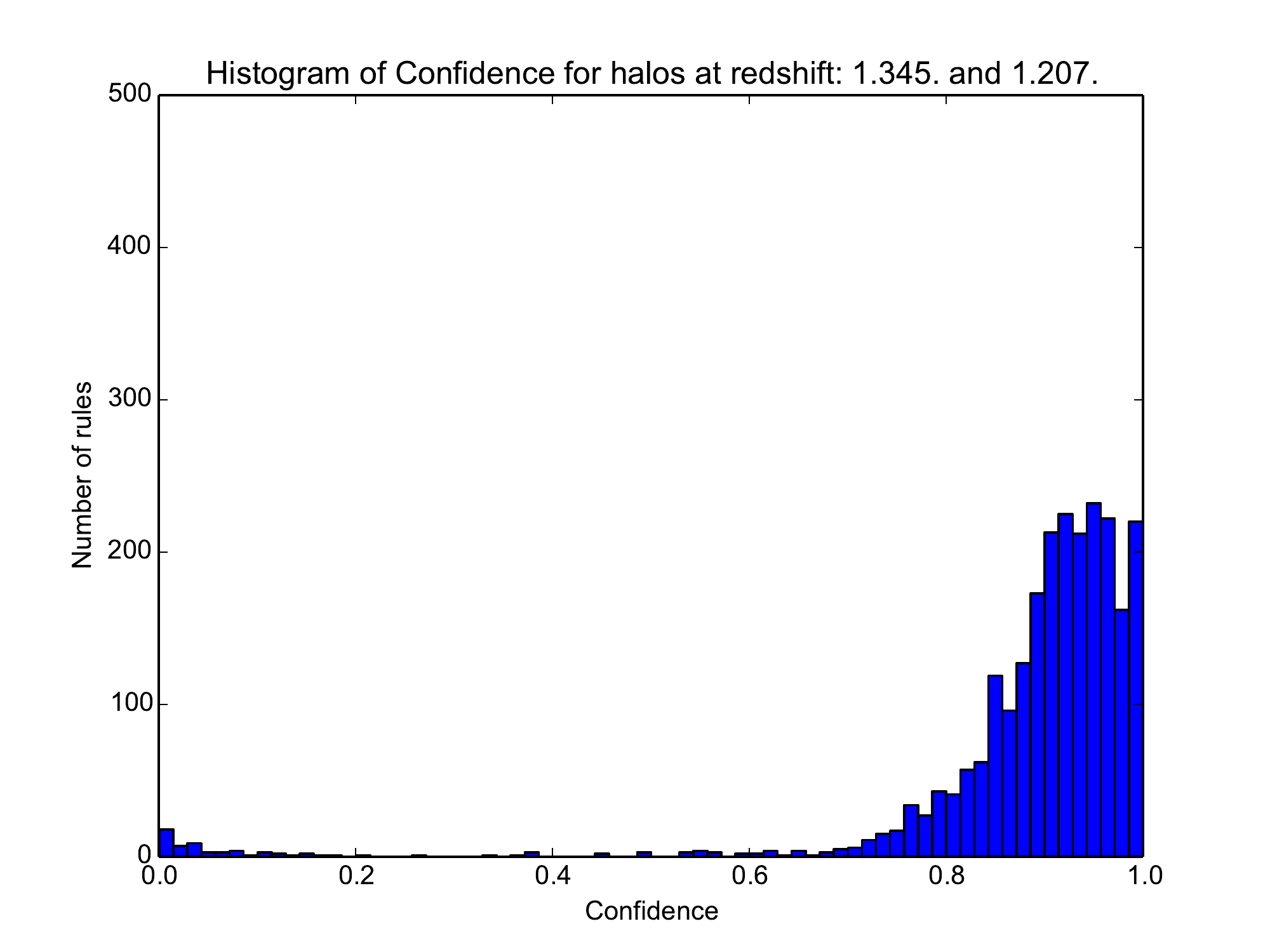}
        \caption[]%
        {{\small}}
    \end{subfigure}
    \caption[ The average and standard deviation of critical parameters ]
    {\small This figure shows the histogram of confidences of rules which correspond to two halos from two different snapshots. For example (a) correspond to snapshots at redshifts z=0.0 (Snapshot II) and z=0.063 (Snapshot I). Rules considered for the histogram would be of the form, \{Some Halo(Snapshot I) $\rightarrow$ Some other Halo(Snapshot II)\}. Peaks around 0.9 present a strong case for hierarchical structure formation.} 
    \label{fig:histplts}
\end{figure*}

\section{Simulation and results}
Dark matter simulations were done using Gadget-2 (see \cite{gadgetpaper}) using $128^3$ particles. Initial conditions were generated using N-Genic. Following values were used for the simulation, $\Omega=0.3,\Omega_\lambda=0.7,\Omega_{baryon}=0.045,\sigma_8=0.83,h=0.7,primodial index = 1, box size=60$ MPc and a smoothing length of $0.013$ MPc. AHF(Amiga Halo finder) (see \cite{ahfpaper}) was used to find halos.  Association rule analysis was applied to the output of AHF to generate merger trees. AHF itself offers a tool to generate merger trees. When compared, both results were identical. 
Hierarchical structure formation says that smaller halos merge to become bigger halos. This can be put to test by plotting a histogram of confidences of rules which corresponds to halos in different snapshots. For example, a rule of form $\text{Halo 1(Snapshot1)} \rightarrow \text{Halo 3 (Snapshot2)}$ (confidence = 0.8), indicates that 80\% of particles in Halo 1 of Snapshot 1 go to Halo 3 of Snapshot 2. Such a histogram provides a measure of how much halos are split during their evolution. This is by no means an exact measure since a lot depends on the halo finders. Figure \ref{fig:histplts} shows such histograms when two snapshots at different redshifts were analysed. The peak at around 0.9 makes a strong case for hierarchical structure formation.

Figure \ref{fig:histplts} is a plot of mass accretion history of the simulation. Masses of halos(M(z)/M(0)) is ploted as a function of redshift(ln(1+z)). Plot compares the results obtained from halo AMIGA halo finder with those obtained using association rule analysis. To get Merger Trees one has to look for rules of the form Halo i(Snapshot m) $\rightarrow $ Halo j(Snapshot n), where redshift corresponding to snapshot m is lower. Rules of this form associate halos at lower redshift to halos of higher redshift. Since halos merge, halos at lower redshifts could have multiple such associations. To build halo merger tree one has to choose one ancestor among such multiple associations. The above plot chooses the ancestor based on the support count. The halo corresponding to rule with highest support count is identified as the ancestor. 

\begin{figure*}[h]
    \centering
    \begin{subfigure}[b]{0.65\textwidth}
        \centering
        \includegraphics[width=\textwidth]{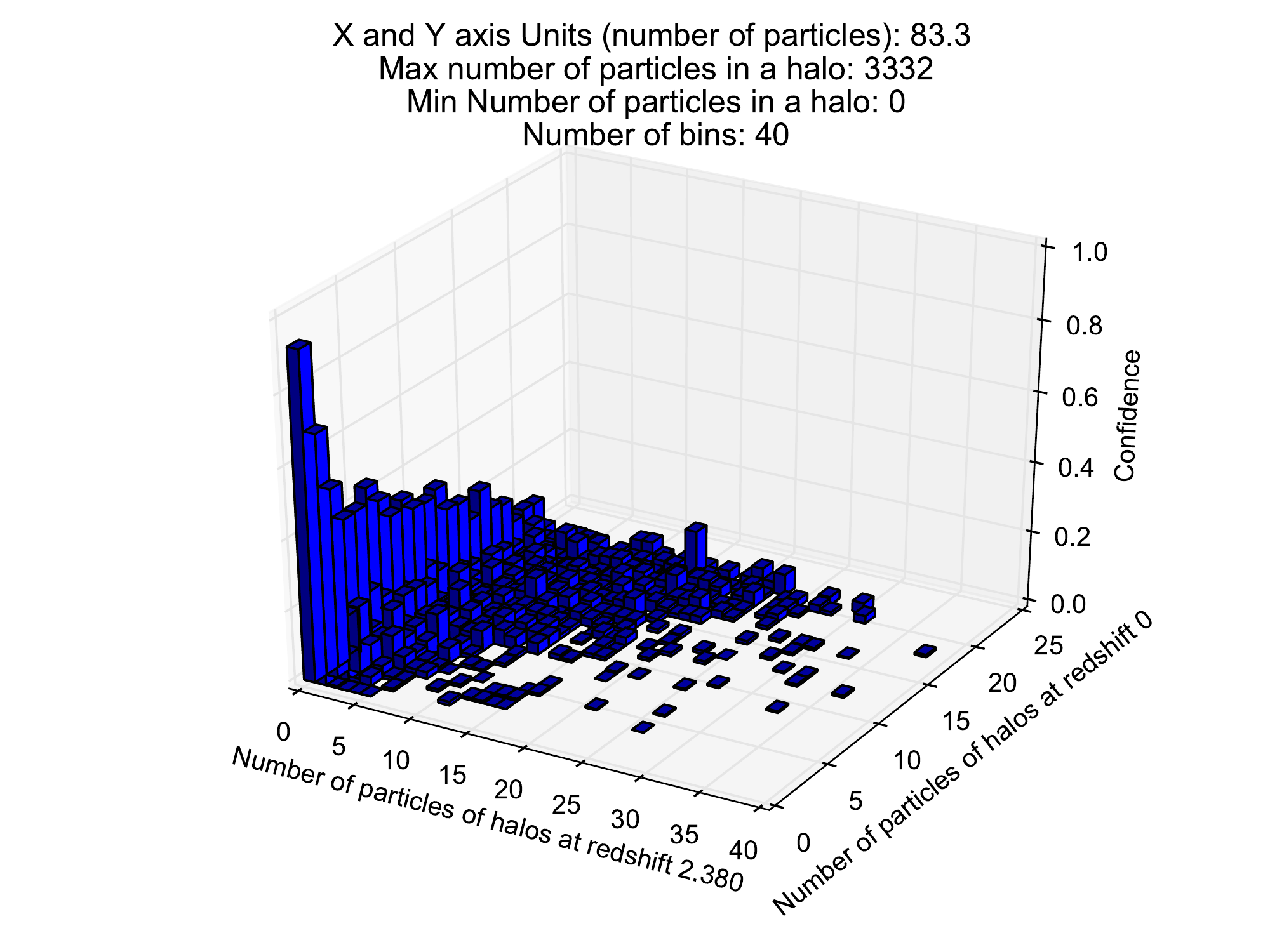}
        \caption[]%
        {{\small}}
    \end{subfigure}
    \caption[ The average and standard deviation of critical parameters ]
    {\small This figure shows confidence of association rules formed using number of particles in halos. Instead of using halo ids to get items, number of particles in halos(discretized into 40 bins) has been used instead. X axis shows number of particles in halos at a redshift of 2.38, Y axis at redshift 0 and Z axis corresponds to the confidence of the rules: $X\text{(number of particles in halos at redshift 2.8)}\rightarrow Y\text{(number of particles in halos at redshift 0)}.$ Peaks in regions where $Y>X$ show that over time halos grow in number.} 
\label{fig:asRuleConf}
\end{figure*}

The above formalism can be used to perform other kinds of analysis as well. One can study how various properties of halos evolve over time. We discretize the physical quantities into small bins and label the halos by not their halo-ids (given by the halo finder) but by the bin they fall into. Now running the association rule analysis on this data will get us rules of the form:
$\text{a}<\text{p}<\text{b(Snapshot2)}\rightarrow \text{c}<\text{p}<\text{d(Snapshot2)}$ . If such a rule has high confidence, it is an indicator that the halos with the given property in range (a,b) evolve over time to halos with the property in range (c,d). Figure \ref{fig:asRuleConf} shows one such plot with number of particles in the halos being the chosen property.

\begin{figure*}[h]
    \centering
    \begin{subfigure}[b]{0.5\textwidth}
        \centering
        \includegraphics[width=\textwidth]{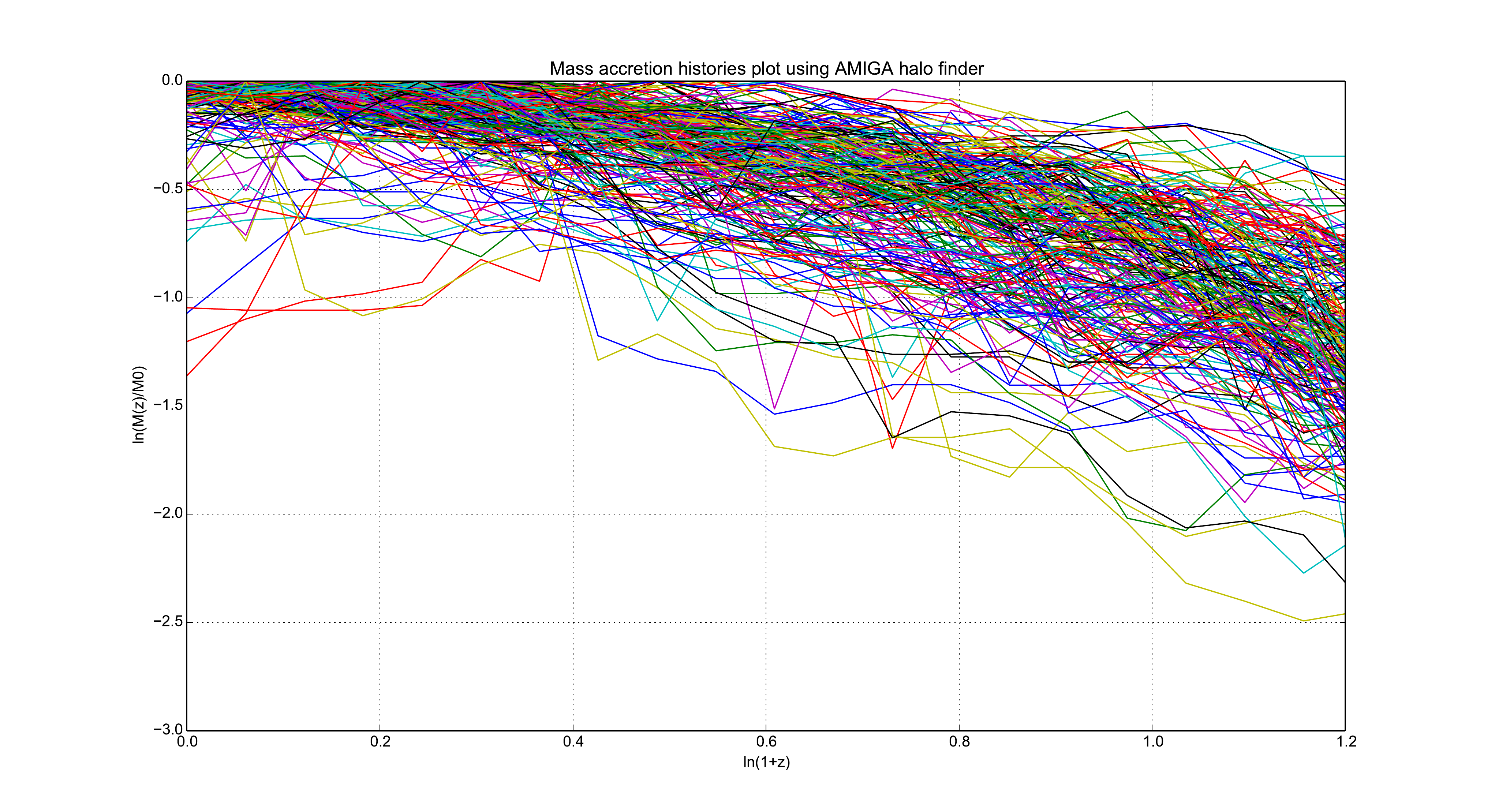}
        \caption[]%
        {{\small}}
    \end{subfigure}
    \hfill
    \begin{subfigure}[b]{0.5\textwidth}  
        \centering 
        \includegraphics[width=\textwidth]{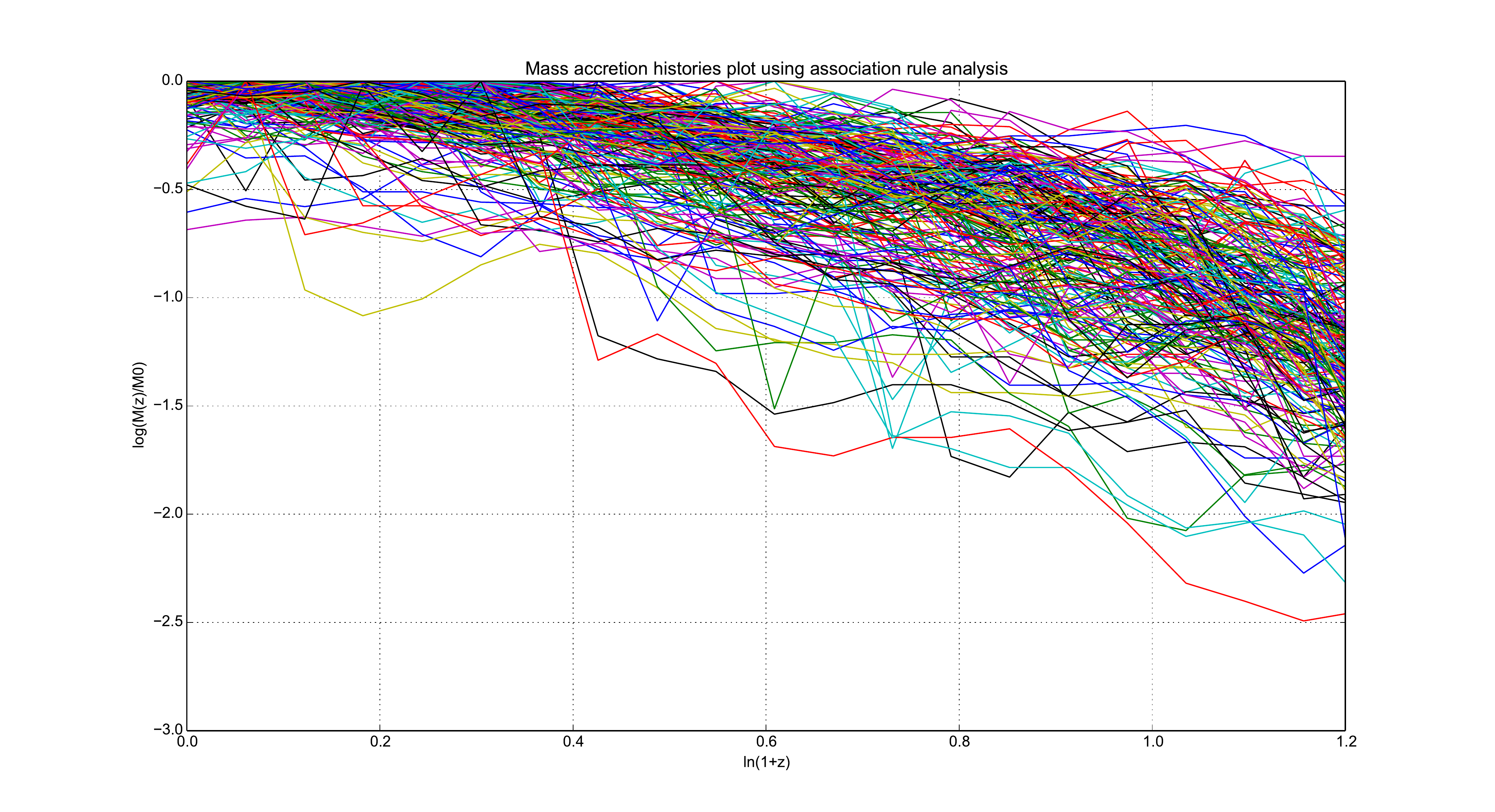}
        \caption[]%
        {{\small}}
    \end{subfigure}
    \vskip\baselineskip
    \begin{subfigure}[b]{0.5\textwidth}   
        \centering 
        \includegraphics[width=\textwidth]{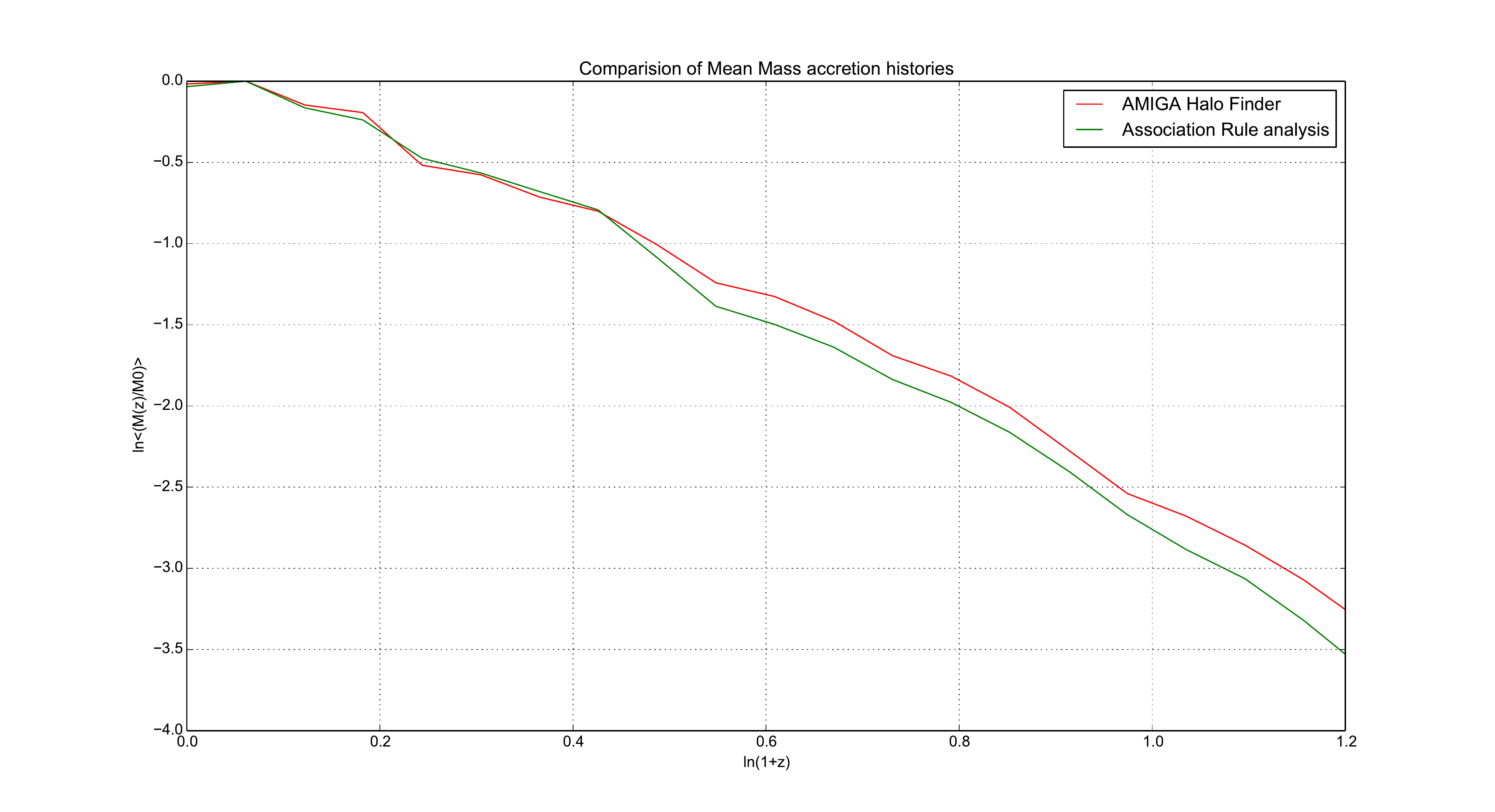}
        \caption[]%
        {{\small}}
    \end{subfigure}
    \caption[ The average and standard deviation of critical parameters ]
    {\small This figure shows the mass accretion histories of the various halos found in the simulation. Halos in the mass range 1 to 3 $\times 10^{12}$ Solar masses are considered for the plot. This translates to around 300 particles. Figure (a) shows M(z)/M0 as a function of ln(1+z), plot using the results from AMIGA halo finder. Figure (b) plots the same using association rule analysis. Figure (c) plots the mean values from plot(a) and plot (b) for comparison.} 
    \label{fig:MAHplts}
\end{figure*}

\section{Conclusion}
Merger trees can be built using association rule analysis. As a by product of the analysis, halo substructure can also be found. By analysing the output, other interesting  relations like how halos are splitting can also be analysed.

\end{document}